\documentclass{vgtc}                          

\ifpdf
  \pdfoutput=1\relax                   
  \usepackage{graphicx}                
  \DeclareGraphicsExtensions{.pdf,.png,.jpg,.jpeg} 
\else
  \usepackage{graphicx}                
  \DeclareGraphicsExtensions{.eps}     
\fi%

\graphicspath{{figs/}{figures/}{pictures/}{images/}{./}} 

\usepackage{microtype}                 
\PassOptionsToPackage{warn}{textcomp}  
\usepackage{textcomp}                  
\usepackage{mathptmx}                  
\usepackage{times}                     
\usepackage{cite}                      
\usepackage{tabu}                      
\usepackage{booktabs}                  
\usepackage{multirow}
\usepackage{url}
\usepackage{amssymb}
\usepackage{amsmath}
\usepackage{float}
\usepackage{wrapfig}
\usepackage{subfigure}
\usepackage{color}
\usepackage{bm}
\usepackage{mathrsfs}
\usepackage{amsmath}
\usepackage[cmintegrals]{newtxmath}

\usepackage[ruled, vlined]{algorithm2e}
\renewcommand{\arraystretch}{1.1} 
\newcommand*{\revise}[1]{\textcolor{black}{#1}}

\usepackage{paralist}
\usepackage{authblk}

\onlineid{7540}

\vgtccategory{Research}

\vgtcinsertpkg

\title{Extending Adjacency Matrices to 3D with Triangles}

\author[1,2]{Rusheng Pan \thanks{e-mail: panrusheng@myhexin.com}}
\author[3]{Helen C. Purchase \thanks{e-mail: helen.purchase@monash.edu}}
\author[3]{Tim Dwyer \thanks{e-mail: tim.dwyer@monash.edu}}
\author[1,4]{Wei Chen \thanks{e-mail: chenvis@zju.edu.cn (corresponding author)}}
\affil[1]{State Key Lab of CAD\&CG, Zhejiang University, China}
\affil[2]{Hithink RoyalFlush Information Network Co., Ltd., China}
\affil[3]{Faculty of Information Technology, Monash University, Australia}
\affil[4]{Laboratory of Art and Archaeology Image (Zhejiang University, China), Ministry of Education, China}

\teaser{
  \vspace{-.7cm}
  \centering
  \includegraphics[width=\linewidth]{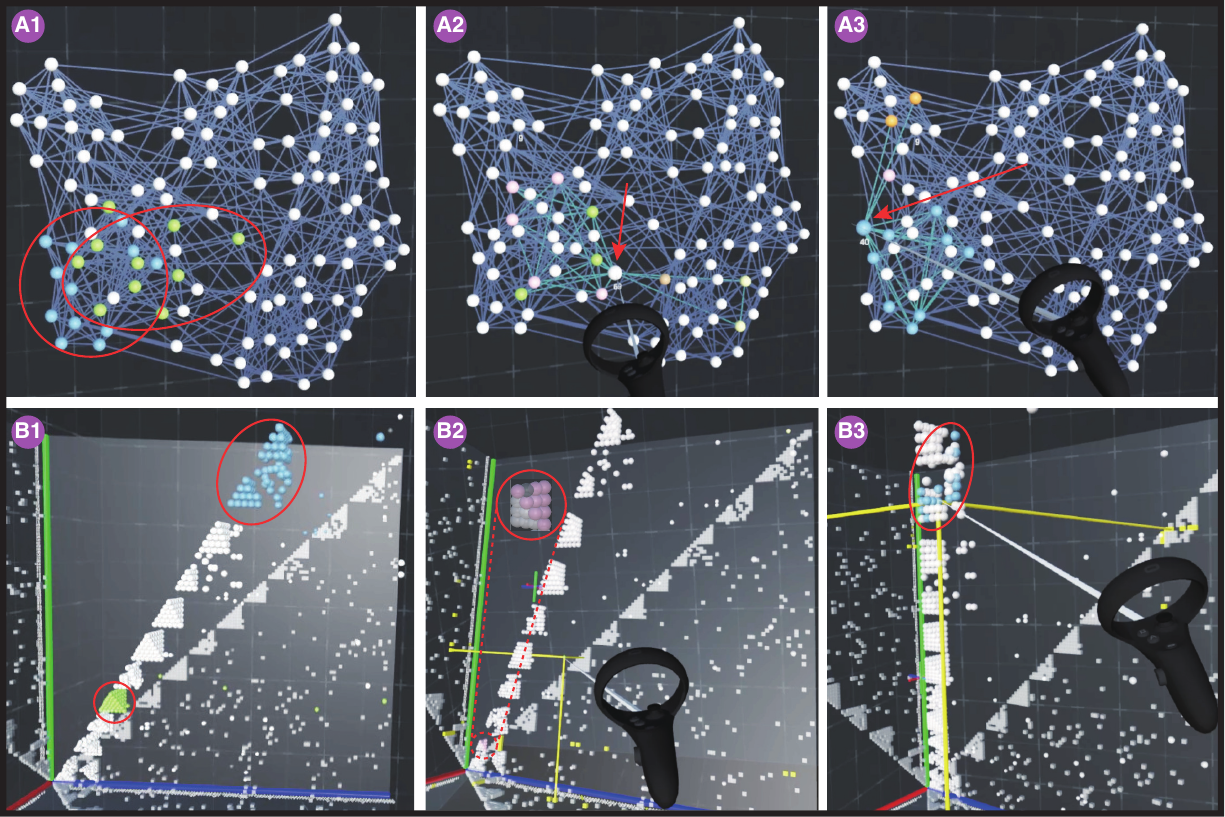}
  \vspace{-.7cm}
  \caption{Three different tasks in the context of traditional node-link (row A) and our 3D adjacency matrix with triangles (row B). A1 and B1: Compare which cluster is denser (green or blue); A2 and B2: Infer which cluster the selected node belongs to; A3 and B3: Estimate the influence~\cite{DBLP:journals/NP/Maksim10} of the selected node. }
  \label{fig:teaser}
}

\abstract{  
Social networks are the fabric of society and the subject of frequent visual analysis. Closed triads represent triangular relationships between three people in a social network and are significant for understanding inherent interconnections and influence within the network. The most common methods for representing social networks (node-link diagrams and adjacency matrices) are not optimal for understanding triangles. We propose extending the adjacency matrix form to 3D for better visualization of network triads. We design a 3D matrix reordering technique and implement an immersive interactive system to assist in visualizing and analyzing closed triads in social networks. The evaluations demonstrate that our method provides substantial added value over node-link diagrams in improving the efficiency and accuracy of manipulating and understanding the social network triads.
} 

\CCScatlist{
  \CCScatTwelve{Human-centered computing}{Visualization}{Visua\-lization techniques}{Graph drawings};
  \CCScatTwelve{Human-centered computing}{Visu\-al\-iza\-tion}{Visu\-al\-iza\-tion app\-lica\-tion domains}{Visual analytics}

}

\begin{document}


\firstsection{Introduction}\label{sec:introduction}
\maketitle

Social networks, such as social media platforms, transaction networks, and cooperative networks, increasingly play a vital role in society. Whilst advances in Artificial Intelligence can provide a wide range of automated tools for analyzing social networks as graphs, visualizations of social networks often fail to adequately support users' comprehension of the data. It is important that the network be presented in a manner that allows users to understand the inherent patterns intuitively and clearly, and to identify connections, clusters, and patterns~\cite{DBLP:journals/ivs/VerspoorOG18, DBLP:journals/bdcc/AkulaG19,DBLP:journals/es/TabassumGACMM23}.
Node-link diagrams are the most popular means of depicting social networks, where nodes represent subjects (people) and links represent the dyadic relations between two subjects. However, node-link diagrams of large-scale networks often contain severe visual clutter (e.g., a ``hairball'' composed of many and dense links), significantly lowering the comprehensibility of the visualization. 
Another common visual network  representation idiom, the adjacency matrix, eliminates the visual clutter problem. An adjacency matrix is a symmetric square matrix where each row and column corresponds to a node and binary values within cells (where rows and columns interact) indicate presence/absence of the links. The adjacency matrix is very good for clarifying edges in high-edge-density graphs with proper row/column reordering, but cannot easily depict some vital sub-structures such as triads (three nodes joined in a triangle). Existing works such as MatLink~\cite{DBLP:conf/interact/HenryF07} and NodeTrix~\cite{DBLP:journals/tvcg/HenryFM07} optimize the visualization of social networks by augmenting the 2D adjacency matrix interactively, but are weak in handling triangle-related tasks. 

The significance of triangles (There are open and closed triads, but the focus here is on ``closed triads") within a social network lies in their ability to facilitate a deeper comprehension of the interconnections and influence between subjects\revise{~\cite{huang2014mining}}. Visualizing the triangles in a social network supports the perception of connections and the completion of multiple interpretation tasks related to triangles. First, unlike edges which represent direct connections between two subjects, triangles depict both direct and indirect (via a common friend) relationships between two subjects. Second, the extent of influence of a subject can be accessed by counting the triangles in which it is involved~\cite{DBLP:journals/NP/Maksim10, DBLP:journals/PhysRevLett/Pastor01 }. Finally, triangles aid in identifying the clusters of social networks in different ways, such as triangle-aware measures and triangle-based motifs~\cite{DBLP:journals/siamcomp/GuptaRS16, DBLP:journals/tkde/RezvaniLLY18, zhang2020community, DBLP:journals/isci/GaoYZ22}. 

Conventional graph representations typically use a two-dimensional (2D) view and rely on flat-screen interactions. Current research provides encouraging evidence that a three-dimensional (3D) visualization of social networks in an immersive Virtual Reality (VR) environment improves users' confidence in observing and understanding the relationships between the subjects~\cite{DBLP:conf/hci/SchroederKA22} over 2D visualization. Specifically, the 3D display of social or communication networks brings practical benefits for visual network comparison \cite{DBLP:journals/tap/WareM08, schneider2016cuboidmatrix, bach2017descriptive, DBLP:conf/cec/ElewahBKRE21, DBLP:journals/tvcg/JoosJSK022}. Most efforts in visualizing graph data in a 3D VR environment focus on extending the node-link diagram with depth information to edges. The problems of overlapping and visual clutter of edges can be alleviated to some degree when edges are separated using the third dimension such that they no longer cross in the same plane. However, the resulting 3D node-link diagram requires rotation interactions to observe the graph with an appropriate view, which reduces the efficiency of exploration and analysis. Moreover, highlighting triangles in a 3D node-link diagram still results in severe visual clutter due to the greater number and increased overlap of triangles when compared to edges. There is not much research on transforming 2D adjacency matrices into 3D space. Bach et al.~\cite{DBLP:conf/chi/BachPF14} design MatrixCube to visualize dynamic networks. They represent the dynamic network at each timestamp as a 2D adjacency matrix and stack the matrices in chronological order. Each 3D cell of the Matrix Cube placed at $(v, w, t)$ represents an edge $e_{vwt}$ between nodes $v$ and $w$ at timestamp $t$. Their later work named Alignment Cubes~\cite{ivanova2017alignment} extends Matrix Cube to support the interactive exploration of multiple ontology alignments. Our work differs from their 3D adjacency matrix display of RDF data or dynamic graphs since we focus on using adjacency matrices to represent graph triangles within static social networks as well as designing the corresponding interactions. We propose extending 2D adjacency matrices into the third dimension such that, for each triangle in the graph composed of nodes $n_i$, $n_j$ and $n_k$, the corresponding binary-valued cell at $(i, j, k)$ in the 3D space is set to True.

This work focuses on social networks with high triangle densities, where visualization for exploring and understanding both global topology and detailed ego-networks can be challenging: 
\begin{itemize}
    \item In general, the number of triangles in a graph is significantly greater than the number of edges. Consequently, the visual clutter caused by triangles overlapping is much more severe than that of edge crossings. Visualizing all the triangles clearly in a social network is nearly impossible using conventional visualization methods. 

    \item From a fine-grained point of view, the patterns of the filtered triangles are complicated. For instance, for each node (or link), all the triangles that contain it would have individual visual patterns. Exploring and analyzing these subgroups of triangles is challenging.
    
    \item Social networks with high triangle densities often contain triangle-spanning hierarchies. \revise{For instance, in a clustered graph where each cluster is composed of nodes, triangles are divided into three types with respect to a cluster~\cite{DBLP:journals/isci/GaoYZ22}. That is, one, or two or three node(s) of a triangle could belong to the same cluster. It is critical to visualize these clustered triangles such that their relationship to the cluster hierarchy is clear.} Designing appropriate representations and layout methods for hierarchical representation is difficult.
\end{itemize}

To address these challenges, we propose visualizing graph triangles as a 3D adjacency matrix, then implement a 3D matrix reordering algorithm based on triangle density, and finally evaluate the effectiveness of the 3D matrix representation on tasks involving triangle-rich graphs.

The rest of this paper is organized as follows. \autoref{sec:related-work} reviews related works from three relevant aspects, and the requirements are analyzed in \autoref{sec:requirements}. Our approach is elaborated in \autoref{sec:methods}, followed by the VR implementation in \autoref{sec:implementation}. A user study has been conducted to evaluate the approach, as reported in \autoref{sec:evaluation}. We reflect on the limitations and future work in \autoref{sec:discussion}. \autoref{sec:conclusion} concludes our work. Two usage scenarios are presented in Appendix A, and additional discussions are presented in Appendix B.

\section{Related Work}\label{sec:related-work}
This section presents the relevant studies on visualizing social networks based on triangles, the application of VR techniques for graph visualization, and matrix reordering methods.

\subsection{Triangle-based Visualization of Social Networks}
In the context of visualizing graph triangles in social networks, the focus has mostly been on extracting triangle-based notions like $k$-truss~\cite{huang2014querying} for probing and analyzing the clique-like structures of triangle-rich complex networks. A truss is a cohesive subgraph defined by the number of triangles supporting an edge. Zhang et al.~\cite{DBLP:conf/icde/ZhangP12} define a triangle $k$-core, which provides an efficient extraction and intuitive visualization of clique-like structures. A $k$-truss based community model is proposed by Huang et al.~\cite{DBLP:conf/sigmod/HuangCQTY14}, which shows structural properties effectively and supports the efficient search of $k$-truss communities. Lan et al.~\cite{lu2023spatial} construct a labor employment space network and extract its characteristics using social network analysis. Their results indicate that a triangular mode structure promotes the evolution of the network. Focusing on modeling complex networks, Seshadhri et al.~\cite{DBLP:journals/pnas/SeshadhriSSG20} mathematically prove that low-rank representations have difficulty capturing both connections and clusters of the triangle-rich graphs. Their study shows that most of the low-dimensional graph embeddings fail to capture the triangle structures. Our work addresses the problem of visualizing triangles in triangle-rich graphs. As a result, we extend the 2D adjacency matrix form (showing the presence of edges) to 3D (showing the presence of triads) so as to provide a clear and intuitive visualization supporting understanding and analysis of triangle-related tasks.

\subsection{3D Graph Visualization in VR}
Benefitting from the commodification and advancements of VR in recent years, research on exploring, interacting with, and visualizing 3D graphs in VR context has increased in popularity~\cite{parker1998visualization, yang2006hierarchical, thomas2014spatial, kwon2015spherical, huang2017gesture, drogemuller2018evaluating, capece2018graphvr, erra2019virtual, pirch2021vrnetzer, kortemeyer2022virtual}. Erra et al.~\cite{erra2019virtual} compare VR systems with traditional mouse-keyboard and joypad configurations, showing that VR enhances user engagement in graph visualization tasks by providing an immersive and enjoyable experience, even though VR techniques are more difficult to learn than traditional ones. This raises the question as to when it is easier to see structure in a 3D VR representation of graph data when compared with a 2D screen layout. The 3D VR technologies support visual comparison tasks in social networks better than 2D ones~\cite{DBLP:journals/tap/WareM08, DBLP:journals/tvcg/JoosJSK022}. The performances of 3D and 2D representation in supporting mental map related tasks are compared by~\cite{DBLP:conf/apvis/KotlarekKMEK0S20}. Their results show that 3D enables more accurate interpretation of the network structures than 2D, while 2D outperforms 3D on spatial-memory tasks. Joos et al.~\cite{DBLP:journals/tvcg/JoosJSK022} develop different encodings for 3D node-link diagrams to visualize two networks using a single representation. Their evaluation results indicate a  potential for visual comparison of networks in VR. Our work builds on these apparent advantages of immersive network visualization, demonstrated for visual comparison tasks in social networks, in designing a 3D adjacency matrix representation and a series of interactions for the immersive perception of social networks.

There is much prior work that focuses on exploring immersive representations and interactions for social networks~\cite{greffard2012immersive, nisiotis2020work, sorger2021egocentric, mallavarapu2022exploring}. VRige (Virtual Reality Immersive Graph Explorer)~\cite{drogemuller2017vrige} is designed to boost the immersive analysis of social network graphs in VR. It provides rich interactions for the visualization, exploration, and tagging of social networks in 3D. NetImmerse~\cite{DBLP:conf/hci/SchroederKA22} is built to visualize networks in VR, providing overview, zooming, and details on-demand, and NetImmerse has been shown to enable its users to gain accurate insights and to enhance the exploration of the multidimensional data of social networks. In our work, we focus on extracting and exploring graph triangles in social networks in VR for efficient identification and comparison of connections, clusters and patterns.

\subsection{Matrix Reordering}
Matrix reordering refers to the technique used to optimize the arrangement of rows and columns in a matrix (e.g., an adjacency matrix or correlation matrix) in order to reveal distinct structural patterns. A brute-force enumeration approach traverses at most $n!$ permutations for $n$ rows/columns, which is impractical for any graph with more than about 10 nodes\revise{~\cite{DBLP:journals/tvcg/KwonKCM23}}. Several adjacency matrix reordering algorithms have been proposed for graph visualization, each using different heuristics, for example, spectral ordering~\cite{DBLP:journals/nla/BarnardPS95, DBLP:conf/icml/DingH04} and hierarchical clustering~\cite{TwoGunnar72, michael1998Cluster, DBLP:conf/ismb/Bar-JosephGJ01}. However, there is no consensus on the most suitable reordering method for all types of graphs\revise{~\cite{DBLP:journals/cgf/BehrischBRSF16, DBLP:journals/tvcg/BeusekomMS22}}. The worth of a reordering algorithm cannot be solely measured by computational efficiency and optimization. An appropriate reordering algorithm for a given graph needs to consider the visual structures inherent in the graph. Chen et al.~\cite{DBLP:journals/tkde/ChenS12} order the matrix rows by identifying similarities between matrix columns in sparse undirected graphs. Specifically, they compute the dense subgraphs based on edge density to yield the corresponding matrix blocks. Our work tackles the problem of visualizing the triangles in triangle-rich social networks. To find a ``good'' order for a 3D adjacency matrix, we consider both the triangle density distribution and the triangle patterns.

\section{Requirements}\label{sec:requirements}
\revise{This work focuses on researching dense social networks with undirected, unweighted edges. To characterize the domain challenges and identify the analytical requirements, we interviewed three domain experts in social science from the local University, their areas of expertise include cultural anthropology, political science, and economics. Focusing on their concerns about revealing triangles, we summarize the following requirements for the visualization of a graph:}
\begin{itemize}
    \item [\textbf{R1:}] \textbf{Show all triangles with both coarse and fine granularity.}
    \item \textbf{R1.1:} \textbf{Showing a compact and clear overview of triangles} is essential. It is undesirable for triangles to occlude one another in the restricted visual space, not least because the node and link information would be hidden by the triangles. 
    \item \textbf{R1.2:} \textbf{Details of each triangle should be provided} during the exploration. The precise set of three nodes and links involved in each triangle should be apparent to support in-depth exploration and analysis of the graph.

    \item [\textbf{R2:}] \textbf{Maintain the hierarchical distribution of the triangles.}
    \item \textbf{R2.1:} \textbf{Depicting the triangles needs to consider their hierarchies} within the graph. For graphs with high triangle densities,  triangles are distributed in different hierarchies, and a hierarchical layout that takes into account these hierarchies reveal the triangles more readily.
    \item \textbf{R2.2:} \textbf{Attaching attributes of nodes or links to triangles} will support specific tasks. For instance, the clusters of some social networks are triangle-aware. By encoding the triangles with information about their node clusters, cluster-related tasks will be more efficient and intuitive to perform.

    \item [\textbf{R3:}] \textbf{Provide the context of the overall graph on demand.}
    \item \textbf{R3.1:} \textbf{The node and link information needs to be displayed} alongside the triangles to support full understanding of the graph. Displaying only the triangles results in a very abstract visualization; accompanying node and link information will allow the user to relate the triangles to a concrete representation of the graph. 
    \item \textbf{R3.2:} \textbf{Extracting subgroups of triangles in situ from the overview} is necessary. Users should be allowed to explore triangles that contain nodes or links of interest. A consistent transition of the interaction when exploring the triangles, nodes and links is essential for maintaining users' mental map.
\end{itemize}

\section{The 3D Adjacency Matrix}\label{sec:methods}
To visualize the triangles in the social networks, we extend the 2D adjacency matrix model (rows and columns) to 3D by adding a third dimension (layer). In a 2D adjacency matrix, the intersection of rows and columns denotes an edge, while in a 3D matrix, the intersection of rows, columns and layers represents a triangle. Like the 2D adjacency matrix, the order of nodes along the axes of the 3D adjacency matrix affects the quality of the visualization. We keep each axis in the same node order and design a 3D matrix reordering algorithm by hierarchically clustering triangles. 

\begin{figure}[h]
    \centering 
    \vspace{-.4cm}
    \includegraphics[width=.8\linewidth]{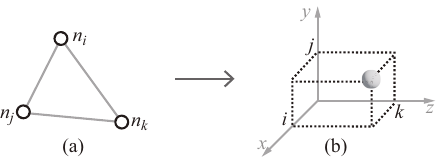}
    \vspace{-.4cm}
    \caption{%
        The transformation from a graph triangle in node-link form (a) to a cell in the 3D adjacency matrix (b).
    }
    \vspace{-.4cm}
    \label{fig:triangle-representation}
\end{figure}

\subsection{Representing triangles in a 3D Adjacency Matrix}
As for 2D adjacency matrices, where the node order in rows and columns is identical, we keep the same node order for layers in the 3D adjacency matrix. The entry in a 3D adjacency matrix (where the $i^{th}$ row, the $j^{th}$ column, and the $k^{th}$ layer intersect) represents a graph triangle composed of the $i^{th}$ node $n_i$, the $j^{th}$ node $n_j$ and the $k^{th}$ node $n_k$, in a triad. \autoref{fig:triangle-representation} illustrates the transformations. 

\begin{figure}[h]
    \centering 
    \includegraphics[width=\columnwidth]{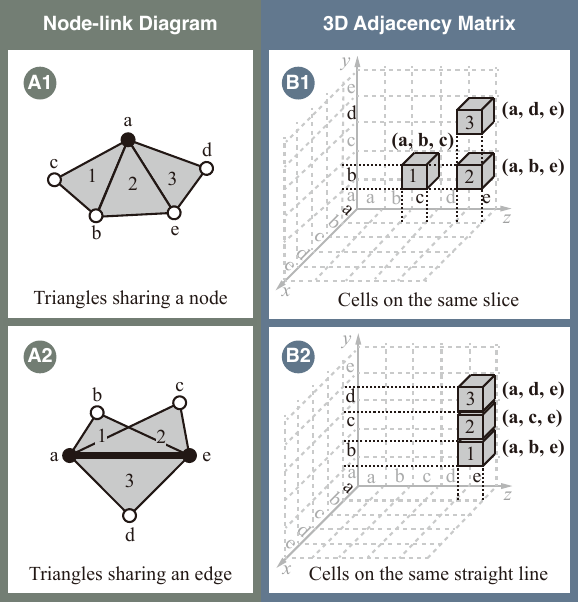}
    \vspace{-.7cm}
    \caption{%
        The comparison of the two triangle patterns with their representations in the node-link diagram and the 3D adjacency matrix. In the 3D adjacency matrix, three triangles sharing a node (A1) are represented as the three cells on the same slice (B1); three triangles sharing an edge (A2) are represented as the three cells on the same straight line (B2). 
    }
    \vspace{-.6cm}
    \label{fig:triangle-pattern}
\end{figure}

In the 2D adjacency matrix of an undirected graph, each edge is usually represented twice, making the node vectors easily readable both horizontally and vertically~\cite{DBLP:journals/cgf/BehrischBRSF16}; this results in a symmetric matrix. Even though the number of edges doubles in this case, there is no visual overlapping, and the topological patterns are easy to identify. In the 3D adjacency matrix, displaying this symmetry would also make it easier to identify the triangles from each of the three axes. As shown in \autoref{fig:triangle-pattern}, triangles containing the same node would be placed on the same slice that is parallel to the coordinate plane of the 3D matrix; and triangles containing the same edge would be placed on the same straight line that is parallel to the coordinate axis of the 3D matrix. However, the number of all the symmetric triangles is six ($3!$) times more than the number of triangles in the graph, and this results in severe visual cluttering and overlapping. We therefore only display the symmetric triangles for nodes that have been selected, and do not display all symmetric triangles. The details can be found in \autoref{sec:interactions}. 

As a result, each triangle composed of the $i^{th}$ node, the $j^{th}$ node and the $k^{th}$ node ($i \textless j \textless k$) appears only once at the coordinate $(i,j,k)$ (unless it is selected, in which case it would also appear at  $(j,k,i)$, $(k,i,j)$, $(j,i,k)$, $(i,k,j)$, $(k,j,i)$ and $(k,i,j)$). The 3D adjacency matrix $A$ of graph $G(V, E)$ is defined as:

\vspace{-.1cm}
\begin{equation} 
    \label{eq:3D-adjacency-matrix}
    A(i,j,k)=\left\{
                \begin{array}{ll}
                  1, \quad i\textless j\textless k \ \ \land \ \ \exists \; \Delta_{ijk} \ \text{in} \ G\\
                  0, \quad \text{else}\\
                \end{array}
              \right.
\end{equation}
\vspace{-.2cm}

\revise{For graphs with node cluster information, we define the cluster of triangles like this: if two or more nodes of a triangle share a cluster XX, then the triangle belongs to cluster XX. Otherwise (three nodes belong to three different clusters), these triangles belong to cluster ``Other'', whose cells are colored in gray.} 

\subsection{Reordering}
In the same way that an appropriate node order needs to be found for 2D adjacency matrices so as to reveal inherent patterns, the order of nodes along the axes in the 3D adjacency matrix affects the quality of the triangle visualization. Our reordering algorithm is based on the principle that triangles within a dense subgraph should be close to one another\revise{~\cite{wang2010triangulation, gupta2014decompositions, samusevich2016local}}. We first transform the 3D adjacency matrix $A$ to a symmetric one $M$:

\vspace{-.5cm}
\begin{equation} 
    \begin{split}
    \label{eq:symmetric-matrix}
    M(i,j,k) = A(i,j,k) \lor A(i,k,j) \lor A(j,i,k) \\ \lor A(j,k,i) \lor A(k,i,j) \lor A(k,j,i).
    \end{split}
\end{equation}
\vspace{-.5cm}

To reorder a graph with $n$ nodes, we view the $n \times n \times n$ 3D matrix as $n$ slices. The $i^{th}$ ($i=1,...,n$) slice is a $n \times n$ matrix equals $M(:,:,i)$, where all the cells represents the triangles containing the $i^{th}$ node. We cluster the slices by their similarities hierarchically, using a strategy based on that used by Chen and Saad~\cite{DBLP:journals/tkde/ChenS12}. While they reorder the columns by edge density, we use the triangle density~\cite{DBLP:journals/siamcomp/GuptaRS16} to reorder the slices. \revise{The similarity matrix $S$ is defined as a 2D adjacency matrix with each element $S(i, j)$ based on the cosine similarity between matrix $M_i$ at the $i^{th}$ slice ($M_i = M(:,:,i)$) and matrix $M_j$ at the $j^{th}$ slice ($M_j = M(:,:,j)$). Here we define the cosine similarity between two 2 $\times$ 2 matrixes $X$ and $Y$ as:}

\vspace{-.3cm}
\revise{
    \begin{equation} \label{eq:cos-def}
        \cos(X, Y) = \dfrac{ \sum_{i, j}(X \odot Y)_{i, j} }{\Vert X \Vert \Vert Y \Vert},
    \end{equation}
}
\vspace{-.3cm}

\revise{where the numerator of the fraction is a scalar that is the sum of the Hadamard product of $X$ and $Y$. Following is the definition of the similarity matrix $S$:}

\vspace{-.2cm}
\revise{
\begin{equation} \label{eq:similarity-matrix}
    S(i,j) = \left\{
        \begin{array}{cll}
            \cos(M_i, M_j)&,& M_i\neq O \land M_j \neq O \\
          1&,&M_i = O \land M_j = O\\
          0&,& \text{else}\\
        \end{array}
      \right.
\end{equation}
}
\vspace{-.3cm}

\revise{We employ the triangle density $\tau$~\cite{DBLP:journals/siamcomp/GuptaRS16} of graph G for reordering slices, whose definition is: }

\vspace{-.2cm}
\begin{equation} \label{eq:tri-density-def}
    \tau(G)=\left\{
            \begin{array}{cll}
              3t(G)/ w(G)&,& w(G) > 0\\
              0&,& w(G) = 0\\
            \end{array}
          \right.
\end{equation}
\vspace{-.2cm}

where $t(G)$ is the number of triangles in $G$, and $w(G)$ is the number of wedges (a two-hop path) in $G$.

\revise{We employ a recursive approach to identify dense triangle sets based on the similarities between each pair of slices. To be specific, we remove edges in ascending order of similarity, until we split the connected graph into two components. Given that we are systematically removing edges one by one from the connected graph $G_S$, once $G_S$ becomes disconnected, it can be reconnected by adding only a single edge. Consequently, $G_S$ cannot consist of more than two components. Such recursive binary splitting continues until the graph achieves the specified triangle density threshold, which is recorded as a dense triangle set. We then rank these sets in descending order of the number of nodes, and output the indices of the nodes in sequence as the new order. Finally, we reorganize the rows, columns and layers by this new order, respectively. The details of the algorithm are shown in \autoref{alg:matrix-reordering}.}

\begin{algorithm}[h]
    \caption{Reorder 3D Adjacency Matrix using Triangle Density}
    \label{alg:matrix-reordering}
    \DontPrintSemicolon
    \LinesNumbered
    \KwIn{undirected connected triangle-rich graph: $G$, \\
          triangle density threshold: $\tau_{min}$}
    \KwOut{ordered array of node indices: $order$}
    \SetKwProg{Fn}{function}{:}{}  
    \SetKwFunction{FStack}{cut\_dense\_triangles}
  
    $G_S$ $\leftarrow$ a graph whose similarity matrix is $S$ defined in \autoref{eq:similarity-matrix}\;
    $T$ $\leftarrow$ an array of tuples $(i,j,S(i,j))$, where $i\textless j$ and $S(i,j)\textgreater 0$, sorted in ascending order of $S(i,j)$ \;
    $D\leftarrow [ \ ]$ \tcp*[l]{\color[HTML]{888888} array of dense triangle sets in $G$}
    \FStack{$G, G_S, T$}\;
    Sort $D$ in descending order of the node count of each subgraph.\;
    Output the nodes indices of all subgraphs in $D$ as $order$.
    \Fn{\revise{\FStack{$G, G_S, T$}}}{
        $n\leftarrow 0$\;
        \While{\rm $n\textless T.length$ and $G_S$ is connected}{
            \revise{Remove edge ($T[n].i, T[n].j$) in graph $G_S$ \;}
            $n \leftarrow n+1$ \;
        }
        \If{\revise{$G_S$ is disconnected}}{
            $G_{S1}$, $G_{S2} \leftarrow$ the two connected components of $G_S$ \;
            $G_1(V_1,E_1) \leftarrow$ the subgraph of $G$ with the same nodes as $G_{S1}$ \;
            \revise{$G_2(V_2, E_2) \leftarrow$ the subgraph of $G$ with the same nodes as $G_{S2}$ \;}
            \If{$\tau({G_1}) \ \geq \tau_{min}$}{
                $D.append(G_1)$ \; 
            }
            \ElseIf{$\left\lVert V_1\right\rVert\textgreater 1$}{
                $T_1 \leftarrow$ the subarray of $T$, where $T_1[k].i \in V_1$ and $T_1[k].j \in V_1$ for $k=0,1,...$
                \FStack{$G_1, G_{S1}, T_1$}
            }
            \If{$\tau({G_2}) \ \geq \tau_{min}$}{
                $D.append(G_2)$ \; 
            }
            \ElseIf{$\left\lVert V_2\right\rVert\textgreater 1$}{
                $T_2 \leftarrow$ the subarray of $T$, where $T_2[k].i \in V_2$ and $T_2[k].j \in V_2$ for $k=0,1,...$
                \FStack{$G_2, G_{S2}, T_2$}
            }
        }
    }
\end{algorithm}

\section{Implementation in VR}\label{sec:implementation}
We implement our 3D adjacency matrix in VR, thus providing a true 3D visualization for complete exploration. Although other alternatives, like a 3D projection in 2D view, are usable, the advantage of our original and novel technique could be limited more or less. \revise{Our 3D adjacency matrix is placed alongside a node-link diagram for reference. During the exploration of the design choices, we have directly rendered the 3D adjacency matrix, and the visualization is hard for most users to follow. In view of the fact that most people are not that used to the 2D adjacency matrix, not to mention the new 3D one. As the most intuitive representation of graph data, a node-link diagram is necessary.}

\subsection{User Exploration}\label{sec:interactions}
\begin{figure}[h]
    \centering 
    \includegraphics[width=.9\linewidth]{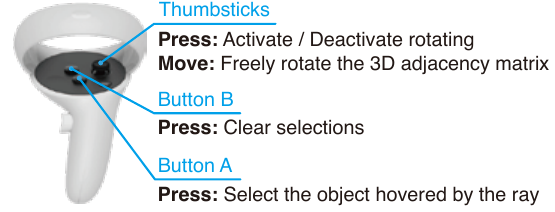}
    \vspace{-.2cm}
    \caption{%
        The interactions provided by the controller. 
    }
    \vspace{-.2cm}
    \label{fig:interactions-of-controller}
\end{figure}

The interactions supported by the controller are illustrated in \autoref{fig:interactions-of-controller}. To begin with, users are allowed to freely rotate the 3D adjacency matrix by moving the thumbstick to find an appropriate angle to observe triangle clusters or different triangle patterns during the 3D exploration. Taking advantage of the head tracking facility, users can zoom into (out of) the view by walking closer (further away). 

\begin{figure}[h]
    \centering 
    \vspace{-.2cm}
    \includegraphics[width=1.0\linewidth]{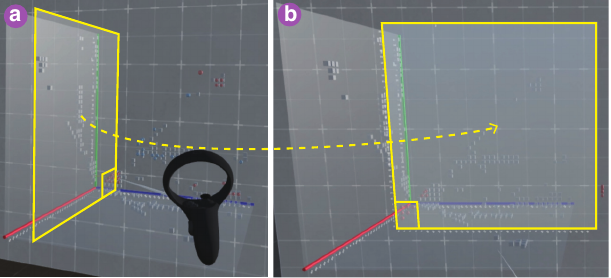}
    \vspace{-.7cm}
    \caption{%
        (a) The 2D adjacency matrix slice colored with a yellow border contains the selected cell, with the intersection of axis $z$ as $w$. (b) All the triangles containing the $w$th node are shown with adding the symmetric triangles (on the left-bottom corner) after rotating the slice to the front of the user.
    }
    \vspace{-.5cm}
    \label{fig:interactions-of-expanding}
\end{figure}

As shown in \autoref{fig:interactions-of-expanding}, the user can select any cell (representing a triangle connecting the nodes $u$, $v$ and $w$) in the 3D adjacency matrix (see \autoref{fig:interactions-of-expanding} (a)). This action will reveal a slice perpendicular to the $z$ axis (the intersection would be (0,0,$w$)); this slice is rotated in front of the user (see \autoref{fig:interactions-of-expanding} (b)). All the triangles containing the $w$th node are plotted on this 2D adjacency matrix slice.

Finally, showing the node-link diagram alongside the 3D adjacency matrix is important for users to select nodes of interest and observe the corresponding triangles in the 3D adjacency matrix. When users select a node in the node-link diagram, all its adjacent links and the supporting triangles (including the nodes) are highlighted, while the other nodes and links are faded. In addition, the relevant triangles are highlighted in the 3D adjacency matrix, and the adjacent links are highlighted as the projection of cells on the axis planes. 

\subsection{Implementation}
\revise{The 3D adjacency matrix of triangle and the VR interactions are developed with the Unity 3D engine, which serves as a popular tool in immersive visualization~\cite{DBLP:conf/avi/LeePPFH06, dai2022robohapalytics}.} The 3D matrix reordering algorithm and all the interactions are programmed in C\#. The force layout of the node-link diagram is yielded from the Python networkx library. All the interactions are implemented based on the Meta Quest 2 Headset and controllers to provide a fully immersive and 3D experience for users' exploration of the 3D matrix.

\section{Evaluation}\label{sec:evaluation}
\revise{We first design two usage scenarios for our 3D adjacency matrix to present the added value of graph triangles and the benefits of visualizing triangles with the 3D matrix. The details are elaborated in Appendix A, which is also available online~\footnote{https://arxiv.org/abs/2306.07588}.}

\revise{Then we conducted an expert user study to evaluate our novel approach to visualizing triangles using a 3D matrix. The complexity of the 3D matrix and the novel conceptualization of how graphs might be visualized meant that it was important that our participants were not visualization novices. As a design choice, we have directly compared the 3D adjacency matrix (Cube) and node-link (NL) in the pilot study, and the result is not good. To emphasize again, since most people are not that used to the 2D adjacency matrix, not to mention the new 3D one. As the most intuitive representation of graph data, NL is necessary. Consequently, we anticipate that the Cube visualization provides substantial added value over NL in improving the efficiency and accuracy of performing triangle-related tasks in social networks.} We compare two conditions: the baseline with the node-link diagram with force layout alone, and the node-link diagram along with a 3D adjacency matrix (NL+Cube). To measure the added value, we used a within-subjects design, where each participant is required to use NL fist and then NL+Cube. Each condition includes three types of tasks: graph-level, cluster-level, and node-level. A questionnaire collected participants' demographic information, ratings for various aspects of the visualization, and qualitative comments. Each participant's task answer and response time were automatically recorded.

\begin{table*}[htbp]
    \centering
    \caption{Tasks for each group. \revise{Meanings of the acronyms are listed as follows: G stands for Group; S denotes Sequence; T represents Task.} }
    \label{table:procedure}
    \renewcommand{\arraystretch}{1.2} 
    \begin{tabular}{@{}c|c|c|cccccc@{}}
    \toprule[1.2pt]
    \multirow{2}{*}{\textbf{Stage}} & \multirow{2}{*}{\textbf{Dataset}}                                         & \multirow{2}{*}{\textbf{Condition}} & \multicolumn{6}{c}{\textbf{Task Sequence for Each Group}}                                                                                                 \\ \cmidrule(l){4-9} 
                           &                                                                  &                            & \multicolumn{1}{c|}{G1} & \multicolumn{1}{c|}{G2} & \multicolumn{1}{c|}{G3} & \multicolumn{1}{c|}{G4} & \multicolumn{1}{c|}{G5} & G6 \\ \midrule
    \multirow{2}{*}{\textbf{Train}} & \multirow{2}{*}{\begin{tabular}[c]{@{}c@{}}\textit{Karate Club} \\ \#Nodes: 34 , \#Edges 78  , \#Triangles: 45\end{tabular}} & NL                         & \multicolumn{1}{c|}{S1 = {[}T1, T2, T3{]}}  & \multicolumn{1}{c|}{S2} & \multicolumn{1}{c|}{S3} & \multicolumn{1}{c|}{S4} & \multicolumn{1}{c|}{S5} & S6 \\ \cmidrule(lr){3-3}
                           &                                                                  & NL+Cube                         & \multicolumn{1}{c|}{S2 = {[}T2, T1, T3{]}}  & \multicolumn{1}{c|}{S3} & \multicolumn{1}{c|}{S4} & \multicolumn{1}{c|}{S5} & \multicolumn{1}{c|}{S6} & S1 \\ \cmidrule(r){1-3}
    \multirow{4}{*}{\textbf{Test}}  & \multirow{2}{*}{\begin{tabular}[c]{@{}c@{}}\textit{American Football} \\ \#Nodes: 115, \#Edges: 613, \#Triangles: 810\end{tabular}} & NL                         & \multicolumn{1}{c|}{S3 = {[}T2, T3, T1{]}}  & \multicolumn{1}{c|}{S4} & \multicolumn{1}{c|}{S5} & \multicolumn{1}{c|}{S6} & \multicolumn{1}{c|}{S1} & S2 \\ \cmidrule(lr){3-3}
                           &                                                                  & NL+Cube                         & \multicolumn{1}{c|}{S4 = {[}T3, T2, T1{]}}  & \multicolumn{1}{c|}{S5} & \multicolumn{1}{c|}{S6} & \multicolumn{1}{c|}{S1} & \multicolumn{1}{c|}{S2} & S3 \\ \cmidrule(lr){2-3}
                           & \multirow{2}{*}{\begin{tabular}[c]{@{}c@{}}\textit{EuAll Email} \\ \#Nodes: 125, \#Edges: 482, \#Triangles: 698\end{tabular}} & NL                         & \multicolumn{1}{c|}{S5 = {[}T3, T1, T2{]}}  & \multicolumn{1}{c|}{S6} & \multicolumn{1}{c|}{S1} & \multicolumn{1}{c|}{S2} & \multicolumn{1}{c|}{S3} & S4 \\ \cmidrule(lr){3-3}
                           &                                                                  & NL+Cube                         & \multicolumn{1}{c|}{S6 = {[}T1, T3, T2{]}}  & \multicolumn{1}{c|}{S1} & \multicolumn{1}{c|}{S2} & \multicolumn{1}{c|}{S3} & \multicolumn{1}{c|}{S4} & S5 \\ \bottomrule[1.2pt]
    \end{tabular}
    \vspace{-.5cm}
\end{table*}

\subsection{Participants}
We recruited 12 participants (P1-P12, average age of 31 ($\sigma=6.51$), 5 females and 7 males) from the local laboratory. They have an average of 4 ($\sigma=2.80$) years of experience in Visual Analysis. Three of them are experienced research fellows with more than eight years of research experience in Immersive Analysis, and the others are graduate students studying Network Visualization or Human-Computer Interaction. 
\subsection{Set up}
To evaluate the added value of our 3D adjacency matrix over the node-link diagram, we designed a 2D node-link diagram in the immersive environment as the baseline, where the encoding of and interactions with the node-link diagram are the same as for the 3D adjacency matrix (except for 3D rotation). The reason for embedding the 2D node-link diagram into a 3D world in VR is to ensure fairness for comparison purposes, especially considering the difference in interaction time, \revise{that is the time to 3D rotate, select by laser and navigate. Otherwise, if we compare the condition of the 2D screen node-link diagram with that of the 3D VR adjacency matrix, the interaction time of the 2D condition has a great advantage over the 3D one, which would make the result unreliable.} Moreover, we chose the 2D layout of node-link diagram instead of the 3D layout, because the 2D layout is quite more typical and commonly used than the 3D layout. 
    
\subsection{Tasks}
\revise{Our work focuses on visualizing graph triangles in triangle-rich social networks. To evaluate the exploration and understanding of the network with both coarse and fine granularity, we consulted domain experts from social science and selected the following classical graph tasks that are closely related to triangles~\cite{DBLP:conf/avi/LeePPFH06, DBLP:journals/inffus/CamachoPOGC20}. Specifically, the three tasks are illustrated from top to bottom.}

\textbf{T1 (R1.1, R2.1).} \textbf{Graph-level: Given a graph visualization with ground-truth clustering and two clusters highlighted, which cluster is denser?} The task focuses on the graph as a whole and requires an effective perception of the densities of different clusters. A cluster is dense if the nodes within the cluster have many connections to one another, and relatively few connections to nodes outside the cluster. The participant needs to compare the two clusters by exploring the one-hop ego network of each node in the node-link diagram and the distribution of triangles in the 3D adjacency matrix, respectively. 

\textbf{T2 (R2.2, R3.1).} \textbf{Cluster-level: Given a graph visualization with ground-truth clustering, clusters shown in different colors but with the cluster for one node not revealed, which cluster does the node belong to?} This task comes from the classical ``cocktail problems''~\cite{DBLP:conf/kdd/SozioG10}, which focuses on identifying a dense subgraph that contains the given query node if the query node is densely connected with this dense subgraph. The participant is required to observe the adjacent links and triangles of the query node through its one-hop ego network, and find a dense subgraph containing it. The dense subgraph cluster is identified as the cluster that the query node belongs to.

\textbf{T3 (R1.2, R3.2).} \textbf{Node-level: Given a graph visualization with two nodes highlighted, which node is more influential?} A node is influential if it is highly connected and its neighbors are also well-connected~\cite{DBLP:journals/NP/Maksim10}. This task is significant for optimizing the allocation of resources and improving communication efficiency in social networks~\cite{DBLP:journals/PhysRevLett/Pastor01}. As a result, the task of finding the most influential spreader node is important. Note that each connection between the neighbors forms a triangle containing the spreader node. Thus, the defined influence of a spreader node can be visually perceived by aggregating the count of its adjacency links and the supporting triangles of it in its one-hop ego network. In this task, the participant is asked to select the most influential node by comparing the related links and triangles.

The participants performed these tasks under the two conditions on three different social network datasets. The details of each dataset are described in \autoref{sec:procedure}. All the graph datasets are undirected, unweighted real-world graphs with ground-truth cluster memberships for the nodes. Each node belongs to exactly one cluster. To avoid introducing bias into the responses to the tasks, we anonymize the graph attributes by replacing the name of the dataset, the node labels and the clusters with random indices. That way, the participants could focus on the topological structures of the graphs themselves (rather than their real-world context).

\subsection{Procedure}\label{sec:procedure}
Before the study, we introduced the research problem to the participants, collected their demographic information, and obtained their consent to observe their behaviors during the study. To reduce the performance variation caused by a learning curve (performance-attempt times, typically a logarithmic trend line)~\cite{DBLP:reference/ml/Perlich17} during the formal tasks, we first asked them to complete the three tasks on a training dataset, a small graph \textit{Karate Club}~\cite{zachary1977information}, containing 34 nodes, 78 edges, and 45 triangles. This social network is composed of social ties among the members from two university karate clubs (colored red and blue). The participants are asked to wear the VR headset and take the controller before training, and they are trained under both conditions. During the training process, we instruct the participants to use the system by demonstrating how to explore the graph to answer the tasks by means of various interactions. The answers to each task would be provided as references during the training. The total number of tasks per participant during the training is 1 dataset $\times$ 2 conditions $\times$ 3 task types $=$ 6.

After the training stage, each participant performs two rounds of tests. Each test comprises the three tasks on a different dataset, first under condition \textbf{C1} and then under condition \textbf{C2}. The first test uses the \textit{American Football}~\cite{girvan2002community} dataset (115 nodes, 613 edges, and 810 triangles), whose nodes represent US college football teams, edges represent the games between two teams, and clusters of nodes indicate which conference the teams belong to. The second one uses the \textit{EuAll Email}~\cite{yin2017local} dataset (125 nodes, 482 edges, and 698 triangles), whose nodes represent members of a European research institution, edges represent communication between institution members, and clusters of nodes indicate the department of each member.

For each dataset, the answers for the same task type under different conditions are different. To avoid the learning effect, the sequence of tasks under each condition was randomized for each participant. The six conditions correspond to six sequences of the three tasks, respectively. To ensure the same difficulty, the answers to all the tasks are completely the same for each participant. To avoid the sequential effects of the tasks and the variance of the participants, we adopt a 6 $\times$ 6 Latin Square to allocate the tasks. Following the rule of Latin Square, the 12 participants are divided into six groups. Each group contains two participants and takes a column of task sequences in the Latin Square. G1 contains P1 and P7, G2 contains P2 and P8, and so on. The detailed task sequences for the six groups are indicated in ~\autoref{table:procedure}. The participant's answers and the time taken to respond to each task were recorded. The total number of responses per participant in the tests is 2 test datasets $\times$ 2 conditions $\times$ 3 task types $=$ 12.

\begin{table}[h]
    \small
    \vspace{-.2cm}
    \centering
    \caption{The questionnaire used to collect subjective feedbacks on our representations, interactions and the system design, respectively.}
    \begin{tabular}{|cl|}
        \hline
        \multicolumn{2}{|c|}{Representation}                                                    \\ \hline
        \multicolumn{1}{|c|}{Q1}  & The representation is easy to understand.                  \\
        \multicolumn{1}{|c|}{Q2}  & The representation displays triangles clearly.             \\
        \multicolumn{1}{|c|}{Q3}  & The representation helps effectively explore triangles. \\ \hline
        \multicolumn{2}{|c|}{Interaction}                                                       \\ \hline
        \multicolumn{1}{|c|}{Q4}  & The interactions are easy to learn.                        \\
        \multicolumn{1}{|c|}{Q5}  & The interactions are easy to use.                          \\
        \multicolumn{1}{|c|}{Q6}  & The interactions are useful.                               \\ \hline
        \multicolumn{2}{|c|}{System Design}                                                     \\ \hline
        \multicolumn{1}{|c|}{Q7}  & The visual design is intuitive.                            \\
        \multicolumn{1}{|c|}{Q8}  & The visual design is good-looking.                         \\
        \multicolumn{1}{|c|}{Q9}  & I am willing to use the design again.                      \\
        \multicolumn{1}{|c|}{Q10} & I am confident in the responses to the tasks.              \\ \hline
    \end{tabular}
    \vspace{-.3cm}
    \label{table:questionnaire}
\end{table}

After the two rounds of tests were finished, the participants provided subjective ratings using a questionnaire designed to collect feedback on our representations, interactions and the system design. Ten questions (as listed in \autoref{table:questionnaire}) on five-point Likert scales were presented, with opinions denoted by  1 - strongly disagree, 2 - disagree, 3 - neutral, 4 - agree, and 5 - strongly agree, respectively. The user's behaviors and feedback (including verbal comments) during the procedure are collected and noted by experimenter observation. The duration of the study for each participant was approximately 50 minutes. Each participant was rewarded with a \$20 E-Gift Card on completion of the experiment.

\begin{figure*}[ht]
    \centering 
    \includegraphics[width=.9\linewidth]{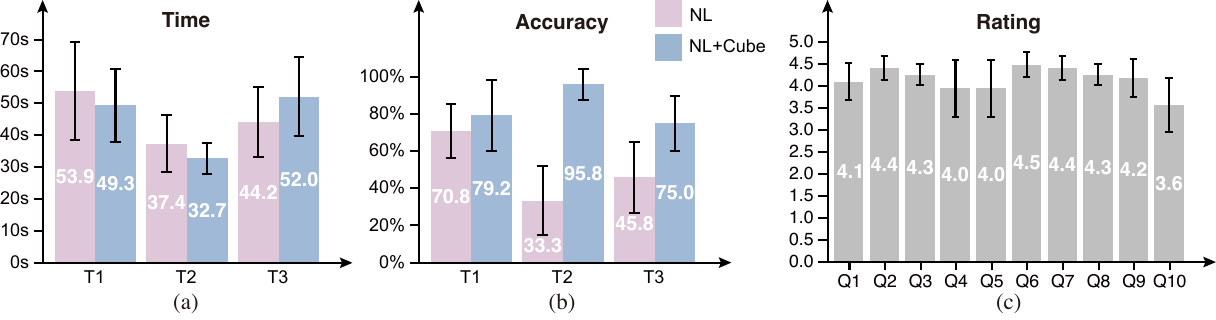}
    \vspace{-.4cm}
    \caption{%
    (a) The average time and the 95\% confidence intervals. (b) The average accuracy cost and the 95\% confidence intervals. (c) The average ratings of the questionnaire and 95\% confidence intervals.
    }
    \vspace{-.4cm}
    \label{fig:user-study-results}
    \vspace{-.1cm}
\end{figure*}

\subsection{Results and Analysis}
We first quantitatively compare the results of the accuracy and time consumption between NL+Cube and NL, before analyzing of the questionnaire, individual feedback and user behavior.

\subsubsection{Task Completion Quality}
We use the paired sample t-test to analyze the differences in response time and accuracy between the two conditions. The p-values of the results are presented in \autoref{table:tasks-ttest}. 
The average performance of the accuracy and time consumption and their 95\% confidence intervals are plotted in \autoref{fig:user-study-results} (a) and (b).

\begin{table}[h]
    \small
    \centering
    \caption{The p-values from the t-tests under two conditions to perform the three tasks. 
    A p-value that is greater than 0.05 means the difference is insignificant. The p-value of the accuracy of \textbf{T3} is less than 0.05, which means a statistically significant difference. The p-value of the accuracy of \textbf{T2} is less than 0.001, which is viewed as a highly statistically significant difference.}
    \begin{tabular}{c|cccc}
        \toprule
        P-value & \textbf{T1} & \textbf{T2} & \textbf{T3}  \\ \midrule
        Accuracy  & 0.501   & \textbf{\textless0.001} & \textbf{0.026} \\
        Time  & 0.645  & 0.377 & 0.361  \\ \bottomrule
    \end{tabular}
    \vspace{-.3cm}
    \label{table:tasks-ttest}
\end{table}

There are no significant differences between the two conditions for response time; we therefore do not need to consider any time-accuracy trade-off and can focus on the accuracy data.

A highly statistically significant difference for \textbf{T2} task accuracy is observed between NL+Cube ($\mu=95.83\text{\%}, \sigma=0.14$) and NL ($\mu=33.33\text{\%}, \sigma=0.33$), with the p-value of \textbf{T2} accuracy equaling 0.000004 (\textless 0.001). The average accuracy of \textbf{T2} using NL+Cube is 62.5\% higher than NL as presented in \autoref{fig:user-study-results} (b). Meanwhile, the p-value of \textbf{T2} response time is 0.377 ($\textgreater 0.05$) indicating no significant difference between NL+Cube ($\mu=32.67\text{s}, \sigma=8.77\text{s}$) and NL ($\mu=37.35\text{s}, \sigma=15.68\text{s}$) , although we note that the the NL+Cube condition takes $4.7$s less on average (see \autoref{fig:user-study-results} (a)). These results suggest that the clear representation of triangles in the 3D adjacency matrix and the use of efficient and effective interactions mean that users are able to identify the ``cocktail problems'' (\textbf{T2}) accurately and quickly.

The p-value of \textbf{T3} accuracy is 0.026 (\textless 0.05), which indicates a statistically significant difference between NL+Cube ($\mu=75.00\%, \sigma=0.26$) and NL ($\mu=45.83\%, \sigma=0.34$). The average accuracy of \textbf{T3} using NL+Cube is 29.2\% higher than NL (see \autoref{fig:user-study-results} (b)). Although the p-value of \textbf{T3} for response time is 0.361 ($\textgreater 0.05$), which indicates no significant difference between NL+Cube ($\mu=52.05\text{s}, \sigma=21.82\text{s}$) and NL ($\mu=44.19\text{s}, \sigma=19.38\text{s}$), we note that (as shown in \autoref{fig:user-study-results} (a)), the NL+Cube condition takes $7.8$s longer on average. In \textbf{T3}, users were required to count the triangles and the links of the two spreader nodes to compare their influence. Even though the 3D adjacency matrix makes it easier to observe the distribution and the number of triangles, it takes longer for users than in NL, where the highlighted triangles are simply too cluttered to count. To sum up, the 3D adjacency matrix significantly improves the accuracy for identifying the details of triangles (\textbf{T3}).

As for \textbf{T1}, the p-values of both accuracy (0.501\textgreater 0.05) and time (0.645\textgreater 0.05) show no significant differences between the two conditions. As illustrated in \autoref{fig:user-study-results} (a) and (b), the accuracy and response time of \textbf{T1} using NL+Cube are both better than in the NL condition, with the accuracy 8.4\% higher and the response time 4.6s less. This result indicates that the 3D adjacency matrix might help with comparing the densities of different clusters (\textbf{T1}) more accurately and efficiently by providing the density distribution of triangles, but no significant differences are detected. 

Overall, the results suggest that the density distribution of triangles (overview) and the context information (detail) are clearly represented in the 3D adjacency matrix, thus improving the task accuracy without efficiency trade off: the time taken to complete the tasks through comparing, identifying and analyzing the visualizations at different levels is similar.

\subsubsection{Ratings of Questionnaire}
Our questionnaire focuses on the use of the NL+Cube condition and covers three aspects: the representation, the interactions and the system design. Our analysis of the quantitative data is supplemented with information from the post-experiment interview and experimenter observations. 

\textbf{Representation (Q1-Q3).} All the participants, apart from P11, agreed that the representation of graph triangles using a 3D adjacency matrix was easy to understand (\textbf{Q1}, 3 strongly agree and 8 agree). P11 has no prior experience with 2D adjacency matrices, and so it was difficult for them to interpret a 3D one. All the participants agreed that the representation displays triangles clearly (\textbf{Q2}, 5 strongly agree and 7 agree) and helps with the exploration and identification of triangles (\textbf{Q3}, 3 strongly agree and 9 agree).

\textbf{Interactions (Q4-Q6).} 10/12 of the participants agreed that the interactions are easy to learn (\textbf{Q4}, 5 strongly agree and 5 agree) and to use (\textbf{Q5}, 5 strongly agree and 5 agree). Participants who disagreed with \textbf{Q4} gave reasons such as issues relating specifically to the VR environment (P4) or the lack of experience in adjacency matrices (P11). However, P11 agreed \textbf{Q5} that the interactions were easy to learn once she had learned the representation after the training. All the participants agreed that interactions are useful (\textbf{Q6}, 6 strongly agree and 6 agree).

\textbf{System design (Q7-Q10).} All the participants agreed that the visual design is intuitive (\textbf{Q7}, 5 strongly agree and 7 agree) and good-looking (\textbf{Q8}, 3 strongly agree and 9 agree). All the participants except for P4 were willing to use the design again (\textbf{Q9}, 4 strongly agree and 7 agree), and P4 explained that this was because they were unused to the VR environment and VR-based interactions. When asked whether they are confident in their responses to the tasks (\textbf{Q10}), 3/4 of the participants agreed (1 strongly agree and 8 agree). The other participants all acknowledged that although they are not entirely confident in the responses with the 3D adjacency matrix, they are more confident than when using NL alone. 

\subsubsection{User Feedback}
At the end, participants were asked to provide written feedback. While some of the comments related to the nature of the immersive environment (e.g. \textit{``the VR glass could be lighter''} (P7)), several participants recognized how the 3D-matrix facilitates a novel and useful way of looking at graphs; for example \textit{``interesting new design and I can see where this additional visualization (3D matrix) is needed.''} (P3) As for all novel visualization methods, some participants found it slightly difficult to learn at first, but they acknowledged that once they had learned how to \textit{``interpret the axes''} (P11), they felt more comfortable. Several participants particularly commented on the usefulness of the 3D matrix for the third task (determining which node has more influence), reinforcing the fact that this triangle-based representation supports important social network analysis tasks. One participant particularly valued this alternative approach to graph visualization: \textit{``I like the representation, I've found it very neat and intuitive. It changes my perspective on network data.''} (P12)

\subsubsection{User Behaviors}
We observed each user and collected their verbal comments and questions during the study. Most participants appeared to relax when they changed from the NL to the NL+Cube condition; we speculate that this was because they felt it would be easier to finish the tasks. As a result, in the NL+Cube condition, they tended to look more at the 3D adjacency matrix for intuitive patterns, and at the node link diagram as a navigation and reference for selecting the nodes involved in the task.

As for the interactions provided in the 3D adjacency matrix, most of the participants used 3D adjacency matrix rotation more often than expanding the slices to quickly identify the triangle patterns for coarse-grained perception. Some of them preferred expanding the slices of their nodes of interest to explore details by analyzing all the triangles containing them. Due to the reason that they focus more on the details, they sometimes lose focus on the tasks and spend more time than the other participants.

\section{Limitations and Future Work}\label{sec:discussion}
\revise{This section discusses the potential limitations of our work and how they may be addressed in future work.}

To begin with, in the current implementation, the user needs to use a controller for precise selection of the objects of interest among a mass of other objects. This is effective for most users, but we acknowledge that replacing the selection interaction from the controller ray with direct hand grabbing and gestures would be easier and more intuitive for users to learn and operate. For instance, users could use specific gestures instead of walking around to zoom in and out of the 3D matrix to freely explore the visualization at various scales. In this work, we first propose the representation of 3D adjacency matrix extended from the standard 2D adjacency matrix approach to show triangles, and we focus on exploring and evaluating the effectiveness of this representation and the relevant necessary interactions. We acknowledge that there might be effective 2D alternatives that don't involve 6DOF navigation in VR. In future work, we would consider designing a greater variety of interactions to improve the user's experience of triangle cells visualization and make more attempts on other 2D alternatives, including 3D projections. \revise{Moreover, we will explore other tasks and domains where our 3D adjacency matrix may provide benefits over the 3D node-link layout, 2D adjacency matrix and other classical graph representations, respectively.}

\revise{The validity of the user study results is somewhat restricted. The participant count (only 12) is relatively low, and most of the participants possessed experience in Visualization, somewhat constraining the generalizability of the findings. The nature of our study makes it challenging to recruit a large number of participants: it takes place in person, and requires participants to learn about both social science and graph visualization, and to adapt to the VR environment. There is no doubt that a larger number and greater diversity of participants would contribute to the generalizability of the findings. In the pilot study, we found that people with no VR experience found it difficult to follow the procedure, since they felt lost and confused. Due to the limitation, we carefully designed the tasks, which require six groups, to minimize the sequence effect as much as possible. As a result, each group had two participants, which should balance the overall performance to some extent.}

Although our method clearly and intuitively visualizes triangles themselves and the triangle density of clusters, it does not have the characteristic of force-directed layout of node-link diagrams where the distances between clusters can suggest their similarity or the strength of connections between them. That is, the spatial distance between clusters in the 3D adjacency matrix in no way indicates their similarity. In our practical hybrid approach, we try to alleviate this shortcoming by showing the node-link diagram is alongside the 3D matrix, such that both triangles and distance information of clusters are conveyed in the same scene. In future work, we will consider the requirement to improve our 3D adjacency matrix reordering algorithm. \revise{Specifically, different ordering algorithms may give rise to visual patterns with varied support for different tasks, beyond those explored in our study, which focused on triangle density in social networks. Just as the reordering of 2D matrices varies based on different tasks or data, there should also be many choices for the 3D matrix. As for the exploration of salient patterns in this new 3D matrix visualization, we included two cases in the usage scenario. More detailed and extensive patterns are still to be explored.}

Last but not least, The evolution of triangles (e.g., the dynamic transition between open triads and closed triads) is important and well researched in social science. This work focuses on visualizing closed triads in static social networks as 3D adjacency matrix cells, which lays the foundation for visualizing open triads and the evolution of open and closed ones using the 3D adjacency matrix in future work, for example, using animation or an adjacent sequence of 3D matrices. Towards a larger space of fields and problems, we expect our work could inspire the broader communities in visualization and VR to explore more extensions from lower dimensions, thus excavating the huge potential of spatial computing.

\section{Conclusion}\label{sec:conclusion}
Focusing on visualizing triangles in social networks, we identify and summarize the challenges of identifying triangle relationships in large dense social networks. This paper presents a new representation by extending the 2D adjacency matrix to a 3D triangle matrix. A 2D cell which denotes an edge evolves to a 3D cell that represents a triangle. The 3D triangle matrix representation is uncluttered and intuitive for exploring and understanding triangle-rich clusters, triangle patterns and one-hop ego networks in social networks. We implement a VR system to provide immersive interaction with the graph triangles. Our evaluations illustrate that visualizing triangles is important and useful when understanding and exploring  social networks. Three different types of tasks focusing on graph-level, cluster-level and node-level are considered to evaluate the effectiveness of our method. Specifically, we compare a Node-Link diagram alone, and Node-Link+Cube hybrid and find that the Cube adds value over Node-Link alone. In the future, we plan to improve the reordering algorithm to adapt to multiple tasks, support hand gesture recognition to enrich our interactions, and generalize our methods to adapt to dynamic social networks.  

\acknowledgments{
The authors would like to thank the reviewers for their valuable feedback. We also wish to thank Shaozhang Dai, Ying Yang, Jim Smiley and Patrick Reuter for their contributions to the paper. This work is supported by the Fundamental Research Funds for the Central Universities (226-2022-00235), and National Natural Science Foundation Key Project (62132017).}

\bibliographystyle{abbrv-doi-hyperref}

\bibliography{template}
\end{document}



\title{Appendix}

\section*{Appendix A: Usage Scenario}\label{sec:usage-scenario}
We describe two usage scenarios of our 3D adjacency matrix to present the added value of graph triangles and the benefits of visualizing triangles with the 3D matrix. The first case focuses on identifying various patterns of triangles in a typical social network, and the second one illustrates how our approach enable the effective exploration of complex social networks by clearly visualizing triangles. For a more intuitive understanding of these immersive analytics cases, please refer to the video demos attached in the supplemental materials.

\subsection*{Identify the triangle patterns}\label{sec:usage-scenario1}
\begin{figure}[h]
  \centering 
  \includegraphics[width=1.0\linewidth]{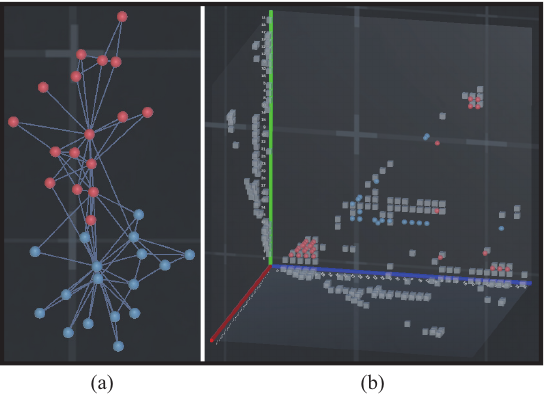}
  \caption{%
  The clusters of Karate Club dataset shown in (a) node-link diagram with a force layout, and (b) 3D adjacency matrix with triangles, respectively. The colors denote the clubs of the members.
  }
  \label{fig:scenario1}
\end{figure}

A male novice of social science wants to explore the \textit{Karate Club}~\cite{zachary1977information} social network, containing 34 nodes, 78 edges, and 45 triangles. This social network is composed of social ties among the members from two university karate clubs (colored in red and blue) collected by Zachary. He is interested in the difference between structures of the two clubs. He first observes the node-link diagram in \autoref{fig:scenario1} (a). Two clusters contain a similar number of nodes and have a similar edge density. It is not that easy to identify the difference between characteristics of the clusters.  

\begin{figure*}[t]
  \centering 
  \includegraphics[width=1.0\linewidth]{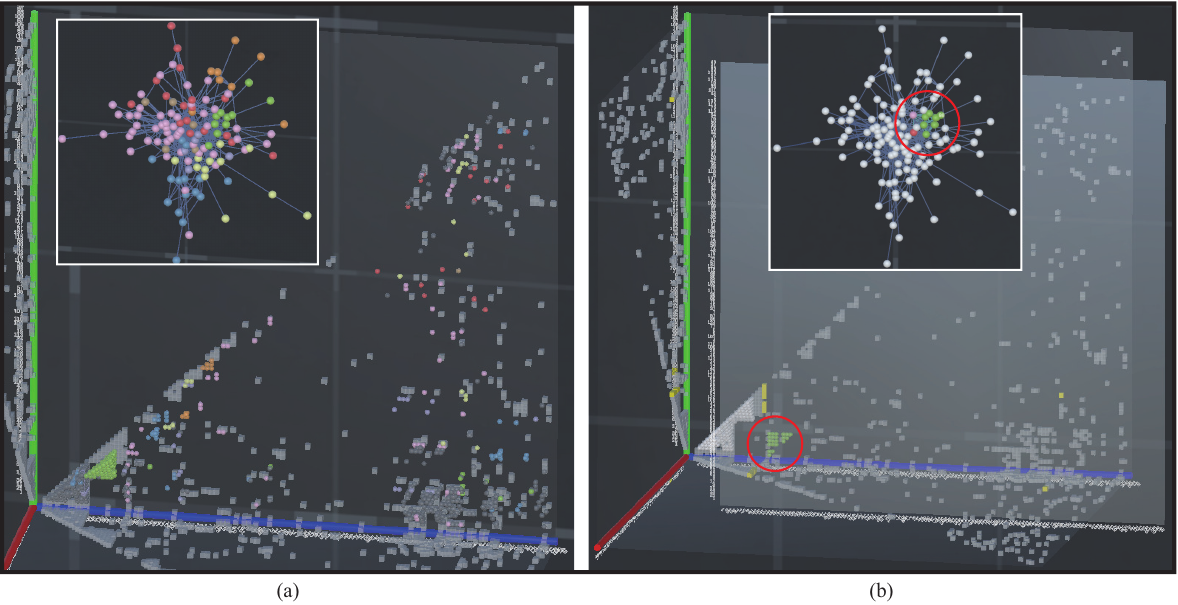}
  \caption{%
  (a) The overview clusters of EuAll Email dataset, where the colors of the cluster (department) could help to understand if the related departments also have related concepts. (b) All but the triangles supporting node 98 are faded.
  }
  \label{fig:scenario2}
\end{figure*}

When he turns to the 3D adjacency matrix in \autoref{fig:scenario1} (b), the two clusters show obviously different characteristics. The triangles within the red cluster are divided into three parts (most on the left bottom, a few on the right bottom, and a few on the right top). The distribution shows that there are three subgroups within the red cluster. The triangles gathering on the left bottom compose the densest subgroup, which locates in the bottom of the red cluster shown in \autoref{fig:scenario1} (a). For the blue cluster, most of its triangles are placed on a straight line that parallels to the $z$ axis (the blue axis). He moves near to zoom in and rotates the 3D matrix to confirm the observation. It means that these triangles share the same edge (or two nodes). To sum up, with the help of our 3D adjacency matrix, he learns that the two clubs are different in their subgroup composition. The red club contains one large and dense subgroup and two small subgroups; while the blue club are basically centered with two central nodes. 

\subsection*{Understanding complex clusters with triangles}\label{sec:usage-scenario2}
Another female user is analyzing a email social network \textit{EuAll Email} (125 nodes, 482 edges, and 698 triangles), whose nodes represent members of a European research institution, edges represent communication between institution members, and clusters of nodes indicate the department of each the member. As shown in the node-link diagram of \autoref{fig:scenario2} (a), the clusters look like a ``hairball'' where the connections and the density of them can hardly be seen clearly. By visualizing the triangles of these clusters, she finds that the green cluster are the densest one, with almost all the green triangle cells gathering together. The pink and blue cluster look dense in the node-link diagram but there are only a few triangle cells of them. The cells distributed in multiple different areas indicates many subgroups exist in the pink and blue cluster. She notices that there are many grey triangle cells which are composed of three nodes from different clusters. These frequent email communications between three or more departments represent that the research institution have many interdisciplinary collaboration or projects. 

Then she focuses on the estimating the influence of some individual members (e.g., node 98). A spreader node is influential if it is highly connected and its one-hop neighbors are also well connected~\cite{DBLP:journals/NP/Maksim10}. It is useful to count triangles supporting the node to identify how many pairs of its neighbours are connected. After she selects the node 98 in the node-link diagram, its one-hop ego network is highlighted (circled in a red border in the node-link diagram of \autoref{fig:scenario2} (b)). However, it is difficult the make out the triangles in the subgraph, because the links are too dense to distinguish. By selecting the slice containing node 98, the slice with all the supporting triangle cells (in the red-border circle at the left bottom of \autoref{fig:scenario2} (b)) rotates to her front. It is much easier for her to estimate the triangles because each cell represents a different triangle. This way, she keeps exploring other nodes and gains an intuitive perception of their strength of influences, which is significant for understanding the communication flow and key spreader nodes of the social network.

\section*{Appendix B: Discussion}
This section summarizes the significance of our contributions and analyzes lessons learned from our trials. 

\subsection*{Significance}
To the best of our knowledge, we are the first to propose a 3D Cube to visualize network triangles. The Cube can be viewed as an extension of the 2D adjacency matrix approach. With an extra dimension, each cell in the 3D adjacency matrix represents a triangle instead of an individual edge, as is the case 2D adjacency matrices. This representation shares the advantage of 2D adjacency matrix views, as compared to node-link diagrams where multiple edges can overlap or cross in the same plane, in that cells are strictly non-overlapping. In a 3-way symmetric representation, triangle cells are so dense that no viewing angle can guarantee that all cells can be seen without occlusion; we therefore hide the symmetric representation of triangles. Symmetric triangles, like the symmetric edges of the 2D adjacency matrix, make it easier to show some topological patterns; the number of symmetric triangle cells is six times the number of actual triangles in the graph and so depicting them makes for a very dense representation.
By contrast, the number of the symmetric edge cells in 2D adjacency matrices is two times the number of edges. 
Consequently, if we remove the symmetric triangles and only display triangle cells in a subspace of the 3D cube such that $x<y<z$, each graph triangle is represented by exactly one cell and the occlusion of triangles is considerably reduced. 
However, without displaying symmetric triangles, some triangle patterns within the graph are less easy to identify. For instance, triangles containing the same node may not be on the same plane. When we only plot triangle cells in the subspace $x<y<z$, the triangles are dispersed on multiple planes. To restore such patterns, we design an interaction of extracting a slice containing a selected node and adding the full set of symmetric triangles associated with that node. Combined with rotation interactions and head tracking, users are able to explore triangles in the 3D adjacency matrix intuitively.

Next, there are various principles which have been applied to compute the reordering of the node rows and columns of 2D adjacency matrices, which may also be transferable to 3D triangle matrices. In a 2D adjacency matrix, each cell represents an edge between the nodes associated with the cell's row and column. Reordering algorithms order both rows and columns symmetrically so only a single axis needs to be considered. Similarly, we apply the 3D reordering algorithm by focusing on one of the axes (row, column or layer). Each axis is the matrix of a node and represents all the triangles supporting that node. As a result, just as when we reorder a 2D adjacency matrix, we are reordering vectors of edges; in the triangle matrix we are reordering matrices of triangles to find a better order of nodes. Since our work focuses on social networks with triangle-sensitive clusters, we employ the triangle density to reorder the nodes. In the resulting arrangement, connected triangle cells are close to one another supporting cluster-related tasks. 

Finally, we evaluated the 3D adjacency matrix by comparing it with the 2D node-link diagrams. Our initial investigations revealed that it was extremely difficult to solve complex tasks using the Cube alone; where therefore compared NL versus NL+Cube to assess the extent of added value provided by the Cube, looking for differences in accuracy and efficiency.  
The interactions and encoding provided for both conditions are the same except that only the Cube supports rotating. This does not introduce any unreasonable bias against the NL condition because there is no need to rotate a 2D node-link diagram. The results show significant differences on two of the three tasks between NL and NL+Cube, where the NL+Cube performs better. The benefits appear to be derived from the clear visualization of triangles in the 3D adjacency matrix, and the useful and efficient interactions. Although no significant differences are detected between the response time  of two conditions. The average time of NL+Cube is less than NL for two of the three. 

\subsection*{Lessons learned}
Initially, we drew triangle cells at every possible place in the 3D adjacency matrix. Consequently, each triangle appeared at six different coordinates. Although the patterns of such triangles are easy to interpret, there are too many redundant triangle cells overlapping the important ones especially for triangle-rich graphs, leading to excessive occlusion. Moreover, displaying all the symmetric triangle cells makes it hard for users to precisely perceive the number of triangles in the graph. As a result, we restrict the space of the 3D matrix to display each graph triangle with only one triangle cell. The optimized result has much less clutter.

Our original implementation contained only the 3D adjacency matrix. Although the triangle cells are easy to interpret, without a node-link diagram, it is difficult to imagine the whole network context of the triangles. The 2D adjacency matrix has a similar restriction, however, as a representation that is more familiar to most people it may be less of a problem. To solve this problem, we combine the familiarity advantage of the node-link diagram  with our 3D adjacency matrix view of triangles. Our experiment  results indicate that this hybrid approach does provide significant added value in different types of tasks.


\bibliographystyle{abbrv-doi}

\bibliography{template}